\documentclass[%
aip,floatfix,
amsmath,amssymb,
 reprint,%
]{revtex4-1}

\usepackage{graphicx}
\usepackage{dcolumn}
\usepackage{bm}

\usepackage{hyperref} 
\hypersetup{
    colorlinks=true,
    linkcolor=blue,
    urlcolor=blue,
    linktoc=all,
    citecolor=blue
           }
\usepackage[utf8]{inputenc}
\usepackage[T1]{fontenc}
\usepackage{mathptmx}
\usepackage{lipsum}
\usepackage{etoolbox}
\usepackage{float}
\usepackage{lipsum}

\makeatletter
\def\@email#1#2{%
 \endgroup
 \patchcmd{\titleblock@produce}
  {\frontmatter@RRAPformat}
  {\frontmatter@RRAPformat{\produce@RRAP{*#1\href{mailto:#2}{#2}}}\frontmatter@RRAPformat}
  {}{}
}%
\makeatother
\begin{document}

\preprint{AIP/123-QED}

\title{High-Resolution Scanning Tunneling Microscope and its Adaptation for Local Thermopower Measurements in 2D Materials}

\author{Jose D. Bermúdez-Perez}
\affiliation{ 
School of Engineering, Science and Technology, Universidad del Rosario, Bogotá 111711, Colombia 
}%
\affiliation{ 
Facultad de Ingeniería y Ciencias Básicas, Universidad Central, Bogotá 110311, Colombia
}%

\author{Edwin Herrera-Vasco}
 \affiliation{ 
Facultad de Ingeniería y Ciencias Básicas, Universidad Central, Bogotá 110311, Colombia
}%
 \affiliation{\mbox{Laboratorio de Bajas Temperaturas, Departamento de Física de la Materia Condensada,} Instituto de Ciencia de Materiales Nicolás Cabrera, Condensed Matter Physics Center (IFIMAC), Facultad de Ciencias Universidad Autónoma de Madrid, 28049 Madrid, Spain
 }%
\affiliation{Departamento de Física Aplicada. Facultad de Ciencias. Universidad Autónoma de Madrid, E-28049 Madrid, Spain
 }%

\author{Javier Casas-Salgado}%
\affiliation{ 
Facultad de Ingeniería y Ciencias Básicas, Universidad Central, Bogotá 110311, Colombia
}%

\author{Hector A. Castelblanco}%
\affiliation{ 
Facultad de Ingeniería y Ciencias Básicas, Universidad Central, Bogotá 110311, Colombia
}%

\author{Karen Vega-Bustos}
\affiliation{
Department of Physics, Universidad de Los Andes, Bogotá 111711, Colombia
}%

\author{Gabriel Cardenas-Chirivi}
\affiliation{
Department of Physics, Universidad de Los Andes, Bogotá 111711, Colombia
}%

\author{Oscar L. Herrera-Sandoval}%

\affiliation{ 
Facultad de Ingeniería y Ciencias Básicas, Universidad Central, Bogotá 110311, Colombia
}%

\author{Hermann Suderow}
\affiliation{\mbox{Laboratorio de Bajas Temperaturas, Departamento de Física de la Materia Condensada,} Instituto de Ciencia de Materiales Nicolás Cabrera, Condensed Matter Physics Center (IFIMAC), Facultad de Ciencias Universidad Autónoma de Madrid, 28049 Madrid, Spain
 }

\author{Paula Giraldo-Gallo}
\affiliation{
Department of Physics, Universidad de Los Andes, Bogotá 111711, Colombia
}%

\author{Jose Augusto Galvis}
\altaffiliation[]{Electronic mail: joseau.galvis@urosario.edu.co}
\affiliation{ 
School of Engineering, Science and Technology, Universidad del Rosario, Bogotá 111711, Colombia 
}%

\date{\today}
\begin{abstract}

We present the design, fabrication and discuss the performance of a new combined high-resolution Scanning Tunneling and Thermopower Microscope (STM/SThEM). We also describe the development of the electronic control, the user interface, the vacuum system, and arrangements to reduce acoustical noise and vibrations. We demonstrate the microscope’s performance with atomic-resolution topographic images of highly oriented pyrolytic graphite (HOPG) and local thermopower measurements in the semimetal $Bi_2Te_3$. Our system offers a tool to investigate the relationship between electronic structure and thermoelectric properties at the nanoscale.

\end{abstract}

\maketitle
\section{introduction}

The invention of the Scanning Tunneling Microscope (STM) by G. Binnig and H. Rohrer brought about a flurry of research in instrumentation\cite{Bining,BiningandDsmith,Bining1981,anirban2022}. Numerous new microscopes were built on the basis of the STM, which enabled imaging of many new properties\cite{BENSTETTER20095100}. The STM still has a prominent role, as it probes the tunneling current between tip and sample, one of the most delicate but also most informative local measurements. For example, the tunneling current can be related to the local electronic density of states, which is useful to study metals, semimetals, semiconductors, and superconductors\cite{Hamann,Tersoff,Crespo2012,GALVIS20121076,ThreeaxisJA,GALVIS20172,Willa2020,BOI2021,Hongyuan,herrera2023}.

In a STM a tip is brought to tunneling distance with a flat sample. When a voltage difference is applied, a tunneling current flows. The tunneling current depends exponentially on the tip-sample distance $d$ as $I_T \sim e^{-2\kappa d}$, where $\kappa$ is the sample workfunction\cite{Hamann,Tersoff,Bardeen}. A change of only one Angstrom in the tip-to-sample distance may produce a change of one order of magnitude in the tunneling current. The tip, typically made of a metal such as gold or tungsten, is mounted on a piezoelectric material that controls its position and can be used to scan the tip over a flat sample's surface.\cite{BiningandDsmith}. There are two main modes for operation of the STM: in the constant current mode, the tunneling current is maintained constant by a feedback circuit, and an image is produced with the z-position of the piezoelement. In the constant height mode, the tip is scanned at a constant height over the surface and the image is made by plotting the current vs position.

A STM can be modified to include new measurement modes for the characterization of physical phenomena not included in conventional techniques, leading to a better understanding of material properties\cite{Lyeo2004,Hwang2022}. For example, one can adapt to measure the thermovoltage, which is related to the conversion of heat into electrical energy. This effect can be quantified through the Seebeck coefficient (S) or thermoelectric power, related to the material figure of merit (ZT) as $(S^2/\sigma\kappa)T$, where $\sigma$  and $\kappa$ are the electrical and thermal conductivity respectively, and T the temperature. A Scanning Thermoelectric Microscope (SThEM) can be adapted from STM to measure S at the local scale, providing a unique technique for the characterization of topographic, electronic, and thermodynamic properties\cite{Lyeo2004,Kim2022}. SThEM has potential in the field of nanoelectronics, where it has been used to investigate the thermoelectric properties of individual nanostructures\cite{Vanis2016}.\\\\
This work presents a high-resolution STM system that operates at room temperature and high vacuum, which has been adapted as a SThEM to incorporate thermovoltage measurements.  In Section \ref{sec:level1}, we discuss the design of the STM, including its key components and features. We then describe the modifications made to the system to enable thermovoltage measurements in Section \ref{STHEM-section}.  In Section \ref{software-section}, we detail the development of the electronic control, acquisition, and data processing software.  In Section \ref{vibrational-section}, we discuss the strategies and instruments implemented to minimize mechanical and electrical noise. Finally, in Section \ref{results-section}, we demonstrate the system's performance by presenting topographic and electronic properties of Highly Ordered Pyrolytic Graphite (HOPG), as well as thermovoltage measurements in $Bi_2Te_3$ single crystal.
\begin{figure*}
\includegraphics[scale=1]{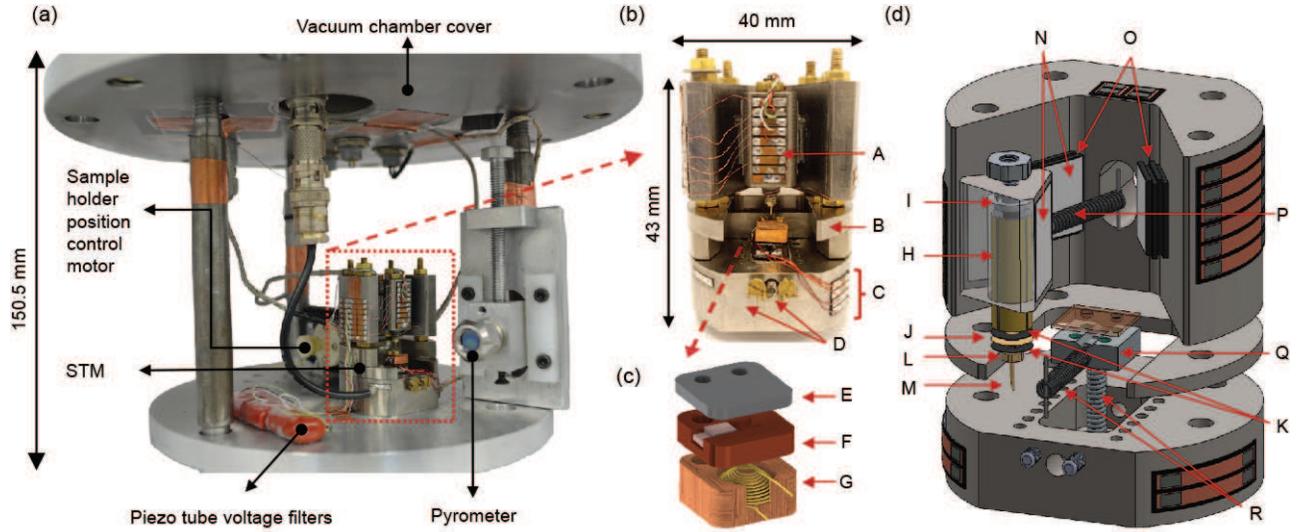}
\caption{\label{fig:CADSTM}(a) Home-built STM system set-up. The system is inserted in a vacuum chamber. Here, the STM and a sample-holder position control motor are firmly fixed. The assembled STM is shown in (b). As a coarse approach system, we use a slider prism (A) and a height control piece (B) to adjust the distance between the slider prism to the sample. The STM bottom part (C) includes the sample holder chamber and the screws (D) to adjust the initial sample position. A detailed view of the sample holder assembly is given in (c). It includes a metal piece (e.g., Chrome, or aluminum) (E) to mount the sample using silver conductive epoxy, a silicon diode (DT-670-SD-1.4H) temperature sensor (F), and a heater (G). In (d), we show the STM's detailed 3D CAD rendering. The piezoelectric tube (H) is attached to the prism by a small Ti piece (I), a copper washer (J) sputtered with gold, provides a ground shield for the current signal, electrical insulation layers (K), the tip holder (L) and the tip (M). The coarse approach system includes alumina plates covered with graphite (N), piezo stacks (O), a spring (P) to attach the slider prism firmly to the STM body, and the sample slider (Q). Finally, the springs are to fix the slider to the sample holder (R).} 
\end{figure*}
\section{\label{sec:level1}Description of the STM \protect}

In Fig \ref{fig:CADSTM}(a), the vacuum chamber cover is shown, which holds the platform with the STM body (Supplementary Fig S1(a)). The design of the STM (Fig. \ref{fig:CADSTM}(b)) is based on previous reports \cite{Pan,Suderow,ThreeaxisJA,Fernandez2021,martin2021}. The STM is made of titanium, which has one of the smallest linear thermal expansions ($8.5 \times 10^{-6}/ ^{\circ} C $) among other metals like stainless steel, chrome, or aluminum, resulting in low structural distortions due to temperature changes\cite{hidnert1943thermal}. This property, combined with the zero magnetic response of titanium, allows for the operation of the STM under extreme conditions of very low temperatures and high magnetic fields\cite{Suderow,ThreeaxisJA,Fernandez2021,martin2021}. These two characteristics are essential for the study of superconducting materials using STM\cite{Crespo2012,GALVIS20121076,GALVIS20172,Willa2020}.


The STM tip is attached to the end of a piezo-tube, which allows for approaching the sample, which is mounted on the sample holder (shown as (E) in Fig.\ref{fig:CADSTM}(c)). In Fig. \ref{fig:CADSTM}(d), the tip (M) is soldered to a small screw with indium, which is then screwed into a small nut (L). To screen the induction from the high voltage signals of the piezo tube electrodes, a grounded copper washer (J) with 10 nm of gold sputtered onto it is used. We also glued two non-conductive epoxies (Stycast 2850FT) washers (K) to provide insulation layers between the tip, the grounded washer, and the piezo tube electrodes. The piezo tube (H) is mounted on the slider prism (A). The slider prism is tightly clamped to the titanium body through a spring (P) and four shear piezoelectrics (O) on each side. The shear piezoelectrics consist of a stack of 4 shear piezoplates glued with epoxy. The stack is 5 mm high and 10 mm wide. The piezo stacks and the prism are joined together using alumina plates (N). These alumina plates are carefully covered with a thin layer of graphite as a lubricant to promote the sliding of the slider prism\cite{Fernandez2021}. This approach system allows the tip movement towards the sample in a maximum range of 10 mm when voltage pulses of a sawtooth signal (up to $60 V_{pp}$) are applied in the piezoelectric shears, producing a global displacement of the piezoelectric tube through a stick-slip motion\cite{Pan,Suderow,ThreeaxisJA,Battisti2018,Fernandez2021}.

\begin{figure}[b]
    \centering
     \includegraphics[scale=1]{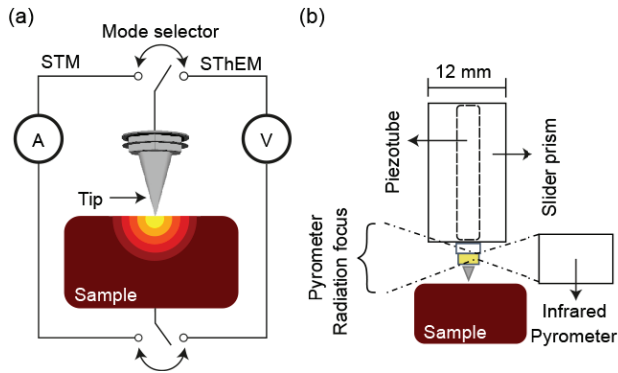}\par
    \caption{\label{SThEM-schema} (a) Schematic representation of the STM/SThEM system showing how to change between tunnel current to thermovoltage measurements. A local temperature gradient is generated in the sample when a tip-sample nanocontact is made (b) Infrared pyrometer schematically configuration, where the focus of the sensor is adjusted and calibrated.}
\end{figure}

\begin{figure*}
\includegraphics[scale=1]{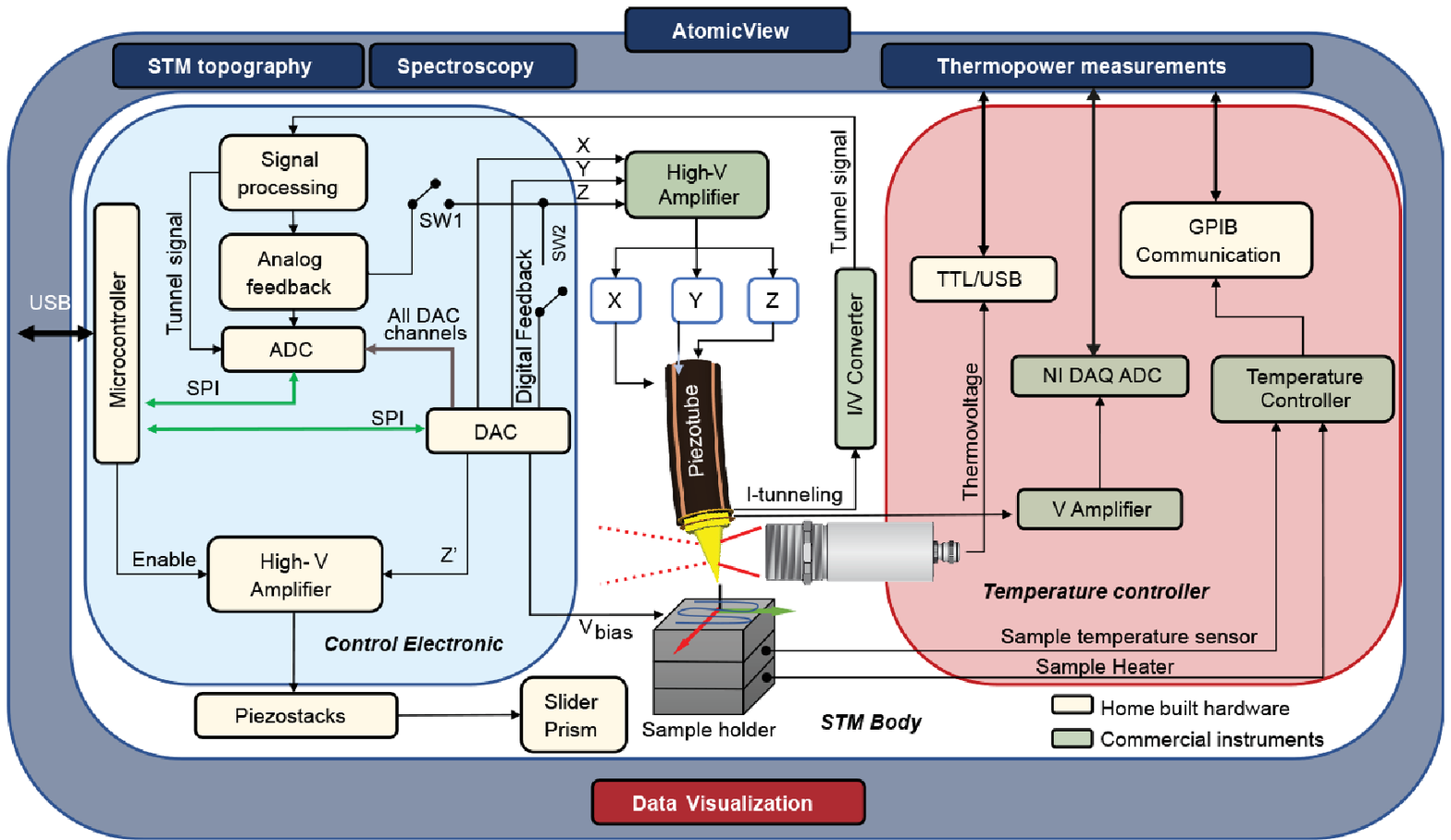}
\caption{\label{fig:blockdiagram}Block diagram of the STM/SThEM electronic hardware set-up, showing the relationship with the AtomicView (AV) user interface (See the section \ref{atomicview-section}). In the dark blue box, the four AtomicView capabilities are shown as "STM topography", "Spectroscopy", "Thermopower measurements" and "Data visualization". In STM operation, The "STM topography" menu allows the STM control parameters (such as Field of view, number of pixels, scanning direction, and time per pixel) to resolve the sample surface in topography mode, and the "Spectroscopy" menu allows the control parameters (as resolution, timer per point and Bias voltage range), which allow the user to obtain sample LDOS information. In SThEM operation, the "Thermopower Measurements" menu is used to measure the Seebeck coefficient. The last AV menu, "Data Visualization" offer visualization feedback as a time function of all the STM/SThEM control parameters. The electronic hardware for STM and SThEM operation are shown in the light blue and red boxes, respectively. The STM/SThEM is shown in the middle of the blank box. Finally, the instrumentation in green boxes shows commercial instruments, and yellow boxes indicate electronic modules, used for the home-built electronic control.}
\end{figure*}

The STM tips are fabricated from Au following the method described in Refs \cite{AGRAIT2003,Rodrigo2004}. Firstly, a cross-section of a 0.008" diameter gold wire is cut using surgical scissors in a glove box (see Section \ref{laboratory}). Secondly, the tip is mounted in the piezo tube. Then, under High Vacuum (HV) conditions, we use the STM to crash the tip in a controlled manner into a clean Au substrate repeatedly until it causes the plastic deformation between the tip and substrate. Repetitive nanocontact indentation between the tip and substrate causes plastic deformation of both, resulting in a narrow connective neck. After several repetitions, the final step involves retracting the tip once quantum transport is established, resulting in the breaking of the neck and the formation of an atomically sharp tip\cite{AGRAIT96}.

The piezo tube is attached to a small titanium piece (marked as "H" in Fig. \ref{fig:CADSTM}(d)) using non-conductive epoxy (Stycast 2850FT), which is then securely fixed inside the slider prism. We use a piezo tube provided by EBL\cite{EBL4}, which has five electrodes: four for X-Y scanning and one for Z-axis elongation. To generate displacement in both the plane and the axis, we polarize the piezo tube by applying a maximum voltage of $\pm$ 125 V. This generates a maximum deformation of 1 $\mu$m in the X-Y plane and an elongation of 350 nm along the Z-axis. The sample is located facing towards the coarse approach system on a metallic sample-holder (E in Fig. \ref{fig:CADSTM}(c)). The sample holder is screwed at a slider (Q in Fig. \ref{fig:CADSTM}(d)), which can be moved back and forth through a sample holder position control motor  (Fig. \ref{fig:CADSTM}(a)) and a system of springs (R in Fig. \ref{fig:CADSTM}(c)) with a minimum step size of 100$\mu$m (the sample-holder position control system is described in detailed in Supplementary Fig S1(b)).

For the electrical connections in the STM, we use vacuum feed-through connectors and low-noise cables. The wires are soldered to the body of the STM via copper tracks, taking care to position it to separate high-voltage signals from tunnel-current signals,  avoiding sources of interference and electrical noise in the more sensitive signal.

\section{\label{STHEM-section}Description of SThEM}

To operate the STM as a SThEM, we have built a temperature control system capable of operating at temperatures ranging from room temperature up to $\sim$ 500 K. The heater consists of a Bakelite piece that has a cylindrical gap with a radius of 4 mm and a height of 3 mm, with 0.4 m of Nichrome wire into the gap in a spiral, which is covered with a thermally conductive epoxy (832TC-A and 832TC-B) (G in Fig. \ref{fig:CADSTM}(c)). We use a silicon diode thermometer glued onto a copper piece with thermal conductive varnish to measure the sample's temperature (F in Fig. \ref{fig:CADSTM}(c)). The sample holder is mechanically attached to a copper piece and the heater using screws (E in Fig. \ref{fig:CADSTM}(c)). To improve thermal conduction, the three pieces are connected through mechanical contact on their flat surfaces, which are coated with thermally conductive epoxy.

To measure the thermopower (S) we use the switch system shown in  fig \ref{SThEM-schema}(a), which allows to change between STM mode to SThEM mode. Once the SThEM mode is activated,  we establish a controlled temperature difference between the tip and the sample by elevating the sample temperature under HV conditions. When the room-temperature tip makes a nanocontact with the heated sample, a temperature gradient appears in the region surrounding the sample-tip contact\cite{Lyeo2004, Blee}. If the sample is a thermoelectric material, this localized $\Delta T$ generates a thermovoltage ($\Delta V$), which is directly related to the S of the sample. After $\Delta V$ reaches its maximum value, the tip is retracted away from the heated sample. S can be found by $- \Delta V/\Delta T$ \cite{Lyeo2004}.

We measure the tip temperature using an HV micro-epsilon\textregistered infrared pyrometer (see section \ref{Sthem-hardware})\cite{Pyrometer}. The infrared pyrometer is located close to the STM using aluminum support, which enables us to adjust the vertical and horizontal position of the sensor's focus (Fig. \ref{fig:CADSTM}(a)). The maximum distance between the sensor and the tip is determined by the tip size and the optical resolution of the infrared thermometer. To prevent measurement errors and temperature contributions from other objects when the SThEM is operating, the pyrometer position is adjusted and calibrated before closing the chamber.

In our case, we set the pyrometer position at 5.4 mm from the slider prism (Fig. \ref{SThEM-schema}(b)) to fill out the field of view of the radiation focus point completely. The emissivity ($\epsilon$) of the tip surface is another relevant parameter for a correct temperature measurement. We calibrate this factor by using a calibrated silicon diode thermometer in three steps: first, we glue a silicon diode sensor (DT-670-SD-1.4H) onto the tip holder. Second, we measure the tip holder temperature when the tip is away from and when it is in contact with the heated sample. Third, we use these silicon diode temperature measurements to modify $\epsilon$ until the pyrometer measuring value corresponds to the diode temperature.

\section{\label{software-section}Description of the STM/SThEM control Hardware and software}

\begin{figure}[b]
    \centering
     \includegraphics[scale=1]{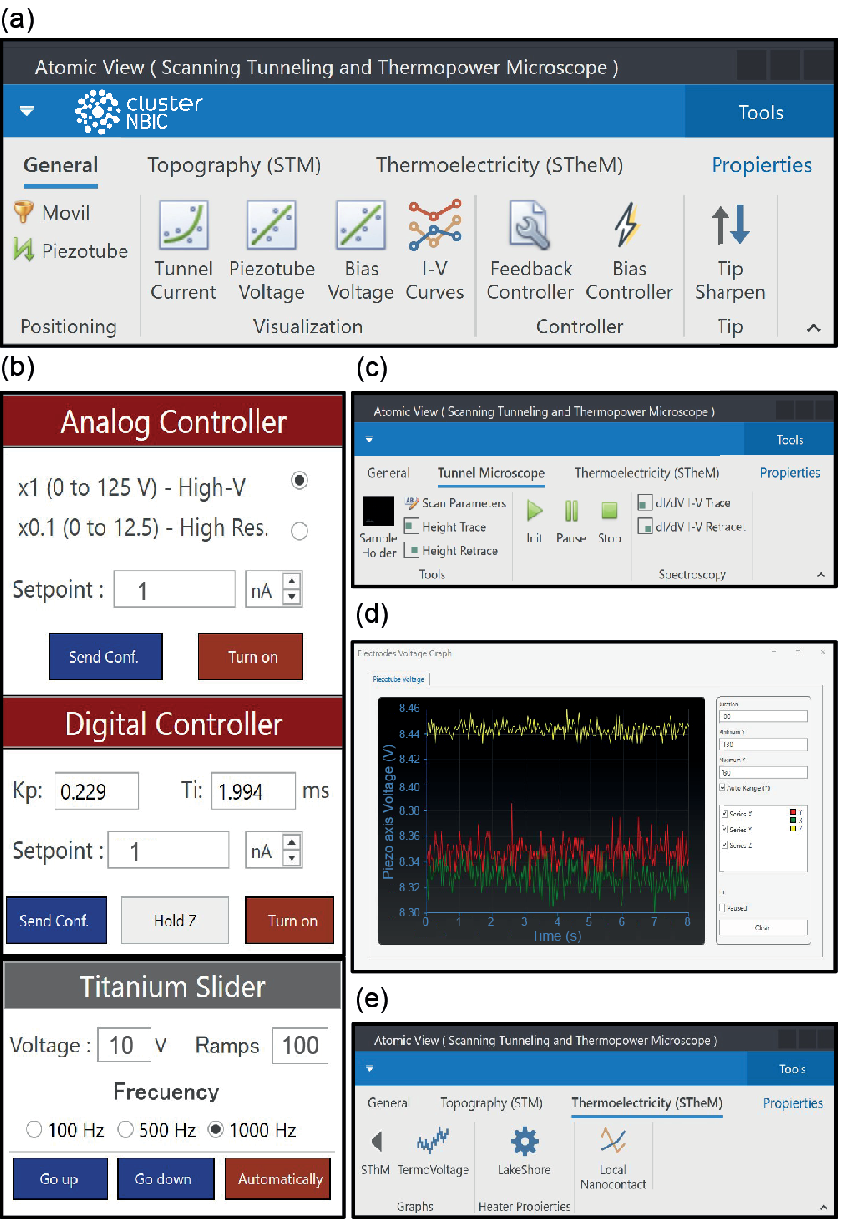}\par
    \caption{\label{atomicview-pages}Screenshots of AV windows. (a) The main interface header shows the different control windows in separate menus. (b) Closed-loop analog and digital current controller window. The analog control parameters are handled externally except for the tunnel current setpoint. The digital part is operated on AV and in the background. (c) STM mode windows. (d) Real-time plot window of the voltages at the piezo tube electrodes. (e) SThEM mode. }
\end{figure}

\begin{figure*}
    \centering
     \includegraphics[scale=1]{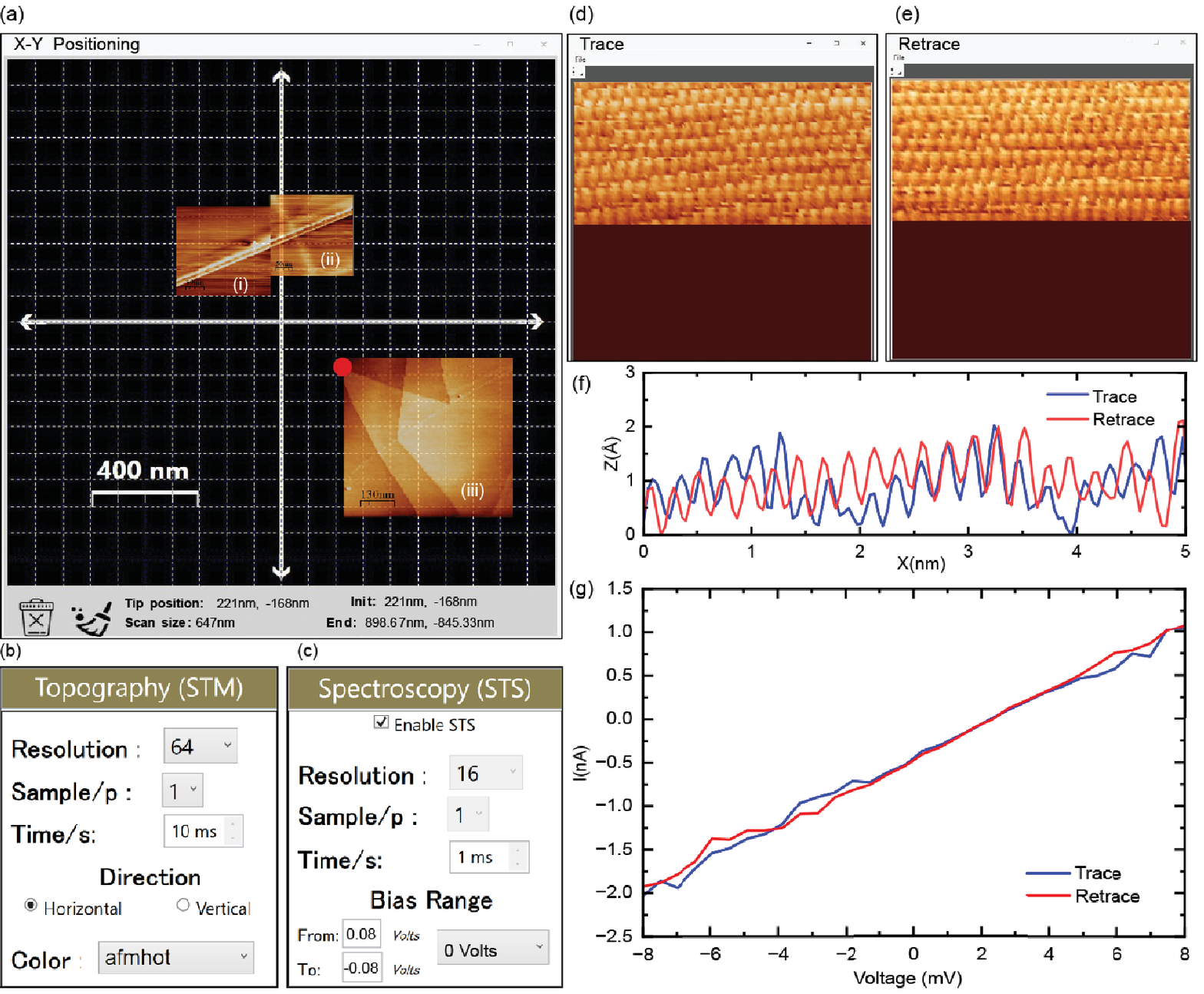}\par
    \caption{\label{screen-inSTM}Screenshots of AV in scanning tunneling mode. (a) X-Y positioning window for topography scanning. The red circle and the white grid mark the tip's position and the piezo tube's maximum scanning area, respectively. We show three different topographies on a HOPG graphite taken at room temperature ($I_T=100 pA$, $V_{bias}=5 mV$) where are evident consecutive atomic layers on the HOPG (i, ii and iii). (b) Control window for topography mode. (c) Control window for spectroscopy mode. (d) Trace and (e) retrace images on atomically resolved HOPG surface, taken with the parameters of (a). (f) Height vs. distance profile and IV Curves in HOPG graphite (g) while the scan in (d) and (e) is measured.}
\end{figure*}
\subsection{\label{STM-hardware}Description of the STM electronic Hardware}

High-quality STM electronic hardware designs are essential for reducing the noise level in the tunnel current\cite{Kamlapure2013}. To achieve this, we designed and built a low-noise electronic system in-house for our STM setup\cite{Jose2022}. The electronic hardware block diagram of the system is shown in Fig. \ref{fig:blockdiagram}. We use a low-noise current trans-impedance amplifier by Femto\textregistered ("I/V converter") to measure the tunnel current, which is linearized and filtered through a "signal processing" module (light blue box in Fig. \ref{fig:blockdiagram}). The current and the analog signals are acquired using an 18-bit Analog to Digital (ADC) chip (ADC8698) with eight separate input channels. The system uses two 16-bit Digital to Analog (DAC) DAC8734 chips, each with four output lines, which provide control voltage signals from $-$5 V to $+$5 V. Six output lines are amplified to a factor of 25 to drive the STM piezoelectric tube, using an ultra-low noise amplifier (TD 250) provided by PiezoDrive\textregistered. For each piezo axis, we use two DACs which are added together, allowing a resolution in the position well below \AA. Furthermore, these output lines have RC filters as close as possible to the piezo tube electrodes to decrease the noise coming from the system power supply. Another output line is amplified to a factor of 12 by a high-voltage amplifier circuit (PA343DF) and is optocoupled into the board to avoid high-frequency interferences. This output line is used to slip the prism (stick-slip motion) applying a sawtooth signal up to +60 V at the piezo stacks (O in Fig. \ref{fig:CADSTM}(d)) (see section \ref{sec:level1}). The final output is used on the sample voltage to establish the tunneling current between the tip and the sample. This output is mounted in a voltage gain configuration to apply Bias voltage signals from $-$10 V to $+$10 V, which directly relates to the energy scale. This voltage range allows us to measure semiconductors with a high band gap such as Boron Nitride (6.08 eV) or low band gaps such as $Ag_2Te$ (0.31 eV)\cite{Cassabois2016,Zhu2019}.

To operate the STM in constant current mode we can switch between analog or digital feedback. The first takes the control signal after the signal processing module (light blue box in Fig. \ref{fig:blockdiagram}). The second operates continuously in the user interface background ("AtomicView") (see section \ref{atomicview-section}). Both systems operate on reading the current value and applying a voltage on the Z electrode of the piezoelectric tube. To control the electronics hardware, we have written a C++ code in an embedded micro-controller (STM32F407VGT6) by ST Electronics\textregistered. We use a standard USB 2.0 to communicate the hardware with the control user interface. The USB is isolated to the main board to avoid noise from the main computer, using a  Digital Isolator (ADUM3160BRWZ) chip. Furthermore, we reduce the whole system noise coming from the 60 Hz line, grounding all the instruments to a separate master ground exclusive for the STM set-up (The whole STM/SthEM system hardware is shown in detail in Supplementary Fig. S2).

\subsection{\label{Sthem-hardware}Description of the SThEM electronic Hardware\protect}

The equipment that we incorporate into the STM hardware to operate it as SThEM is shown in the pink box in Fig. \ref{fig:blockdiagram}. We use an infrared pyrometer thermometer (CT model SF22-C3V) with a temperature range from 223 K to 1248 K and 1 K accuracy to measure the tip temperature. To measure the temperature of the tip in the 1 mm range of the tip holder (L in Fig. \ref{fig:CADSTM}(d)), we add to the pyrometer a close-focus optics lens (TM-APLCF-CT). The specifications of this lens determine the optical path of the sensor, achieving a 0.5 mm focus spot size at 10 mm distance. The sensor transmits data by TTL/USB standard output, used to acquire the tip temperature.

The sample temperature control operates through a commercial Controller (Lake Shore\textregistered CRYOTRONICS 336). One output provides up to 100 W and connects to a 16 $\Omega$ Nichrome heater (G in Fig. \ref{fig:CADSTM}(c)). To measure the sample temperature we use a silicon diode (DT-670-SD-1.4H) anchored to the cooper piece of the sample holder (F in Fig. \ref{fig:CADSTM}(c)) which is connected to one sensor input of the temperature controller. To obtain the thermovoltage when the tip makes a nanoscale contact with the heated sample, we use a commercial low-noise preamplifier (SR560) by Stanford Research. The tip and the sample are connected through a BNC wire at the DC-coupled low-noise differential input signals of 1 to 50000 gain. We acquire the output through a filtered line and using a NI USB-6341 National Instruments data acquisition system (Supplementary Fig. S2).

\subsection{\label{atomicview-section}Description of the user control interface}

The STM/SThEM electronic hardware described in the previous section is handy and can be easily connected to different USB ports on the same computer, allowing remote operation through a user application. We have developed a Windows desktop application (AV) in VisualStudio\cite{Jose2022b}. The AV user software includes all the routines needed to operate the STM/SThEM system, using C as the programming language on a Windows Presentation Foundation (WPF) platform. We communicate with all the electronic hardware via USB standard using AV.

Fig. \ref{fig:blockdiagram} shows a block diagram of the user software modules (dark blue box) and how it is linked to the electronic hardware (light blue, pink, and white box). The software interacts with the hardware through the main page ("AVMain"), which includes four additional menus. The "General" menu (Fig. \ref{atomicview-pages}(a)) is accessed from the main controls and provides options for manual piezo tube movements, operations related to the bias voltage, closed-loop feedback settings, and the operation of the coarse approach system (Fig. \ref{atomicview-pages}(b)). The "Topography STM" menu (shown in Fig. \ref{atomicview-pages}(c)) enables the operation of the STM in topography mode. For feedback of this mode, AV shows the voltage applied on each piezo axis as curves as a function of time (Fig. \ref{atomicview-pages}(d)). In the following sections, we will provide a detailed explanation of this menu along with an example of STM operation. The "Thermopower SThEM" menu (Fig. \ref{atomicview-pages}(e)) is used to operate the STM as SThEM. The "SThEM/Graphs" display shows the tip and sample temperature measurements as a function of time, communicated through GPIB and TTL/USB communication (see the section \ref{STM-hardware}). The "LakeShore/Heater Properties" menu is used to control the sample temperature setpoint, limit the current of the silicon sensor, and the PID operations. The tip-sample nanocontact can be made through piezo tube movement in the "LocalNanocontac/Tools" option or using the coarse approach system in the "General" menu. The last menu "Properties" is used to control the tunnel's current acquisition and processing settings.

In Fig. \ref{atomicview-pages}(b), we show the "Feedback Controller" window in AV, where we can set the control parameters settings, switch between digital and analog controller, and slip the slider prism (A in Fig. \ref{fig:CADSTM}(b)). Another AV routine captures the tunnel current and the voltage in the piezoelectric electrodes through an ADC (see section \ref{STM-hardware}) chip. These values are stored and plotted on an "Electrode voltage" window on the page "General" (Fig. \ref{atomicview-pages}(d)) as a time function. AV also uses several DAC outputs through USB standard to the electronic hardware (see section \ref{STM-hardware}). One output is to control the bias voltage between the sample and tip. AV also employs other outputs to apply voltage to the piezo tube electrodes. During the scanning routine, it is possible to manually or automatically stop the closed-loop controller at each pixel and perform spectroscopy measurements.

\begin{figure}[h]
    \centering
     \includegraphics[scale=1]{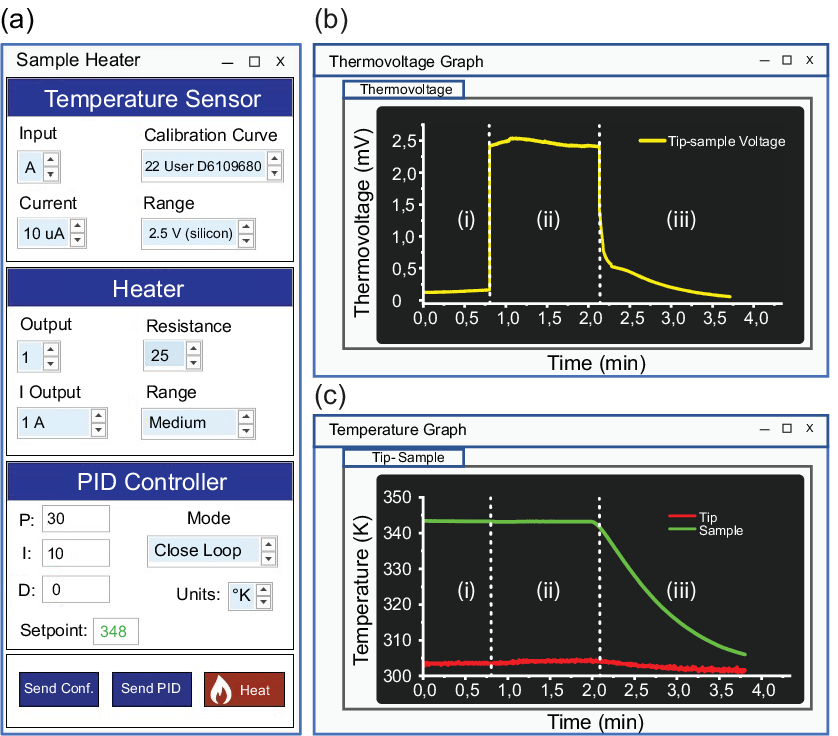}\par
    \caption{\label{screen-inSThem}Screenshots of AV in thermopower mode. (a) Thermovoltage graph window. We show time profiles of the measured voltage in $Bi_2Te_3$ when at room temperature tip make a nanoscale contact with the sample at 343 K. The tip and sample temperature profiles are shown in (b) as a function of the time. In Both graphs we can identify three regions; (i) the tip is away from the sample, (ii) the tip makes a nanoscale contact, and (iii) the sample heater is turned off with the tip in nanoscale contact with the sample. (c) Sample heater setting window.}
\end{figure}
   %
    
\begin{figure*}
    \centering
     \includegraphics[scale=1]{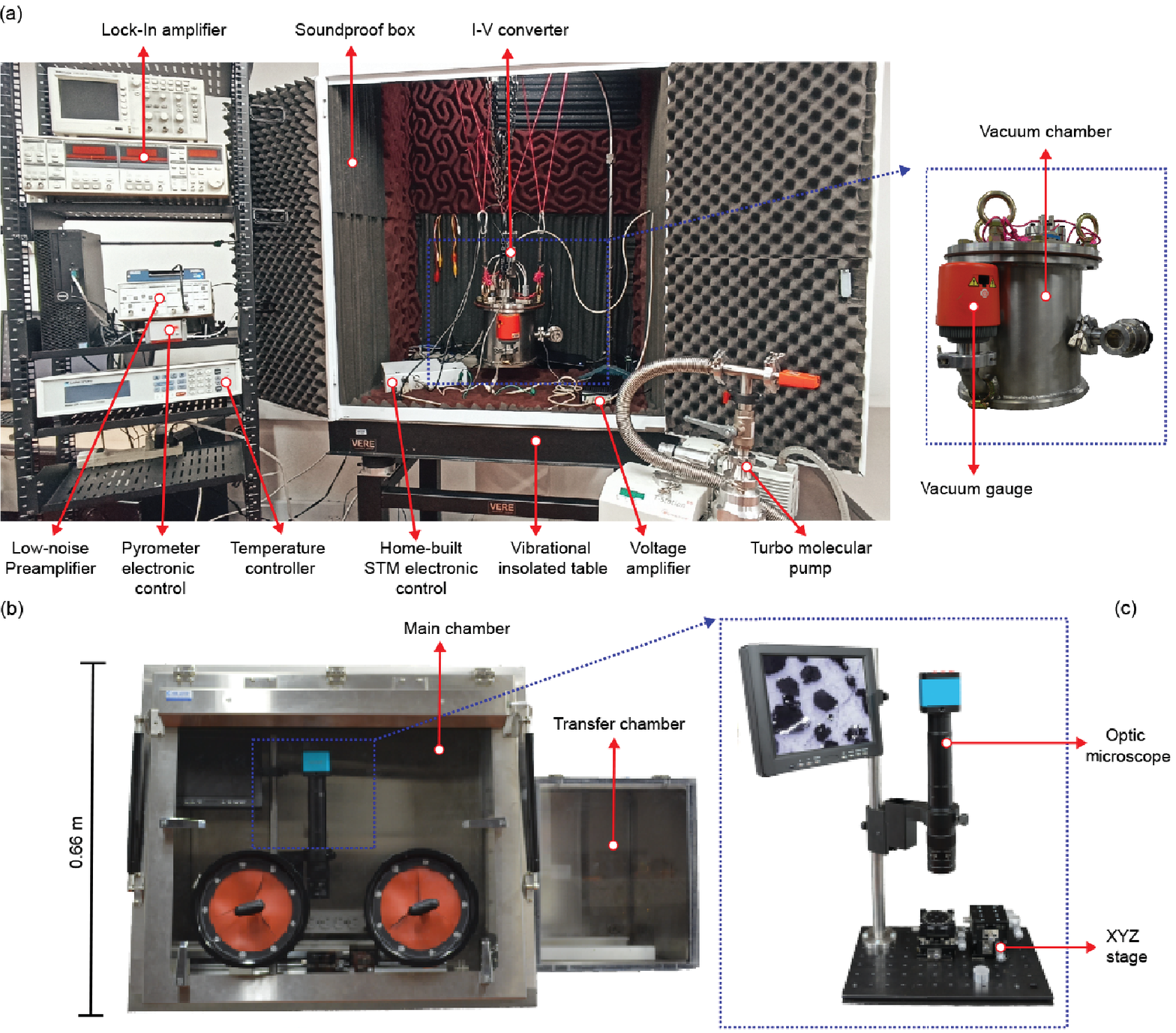}\par
     \caption{\label{laboratory}(a) STM/SThEM set-up. (b) Inert-controlled environmental Stainless steel glove box. (c) Deterministic layer-transfer exfoliation stage.}
\end{figure*}

In Fig. \ref{screen-inSTM}, we present screenshots to illustrate a typical STM topography operation mode. The measurements were taken at room temperature using a gold tip and a HOPG sample. In Fig. \ref{screen-inSTM}(a), we show the X-Y positioning window used for topography operation mode. We can select the size and position of the scanning window, covering areas from 1 nm$^2$ to $2 \mu$m$^2$. The images (i), (ii), and (iii) show topographies in different places of the same scanning window, where some HOPG steps are easily identifiable as atomic layer steps between consecutive graphite planes (more details in Fig. \ref{screen-inSTM}). The small red circle in (iii) indicates the actual tip position in real time. The image settings, such as resolution and time per pixel, can be configured in the "Topography (STM)" menu (Fig. \ref{screen-inSTM}(b)). Furthermore, the "Spectroscopy (STS)" menu controls the parameters for the STM spectroscopy mode (Fig. \ref{screen-inSTM}(c)). Fig. \ref{screen-inSTM}(d) and (e) present a topography in a field of view (FOV) of some tens of nm, where the HOPG atomic lattice is visible (red and blue lines in Fig. \ref{screen-inSTM}(f)) and its metallic behavior in the I-V curves (red and blue lines in Fig. \ref{screen-inSTM}(g)). The data and image analysis were done using WSXM as processing software\cite{Horcas2007}.

In Fig. \ref{screen-inSThem}, we show AV examples screenshots for the SThEM configuration mode. In these measurements, we use a commercial platinum-iridium (805.ASY.STM) tip and a $Bi_2Te_3$ sample. We measure the temperature and the tip-sample voltage as a function of time in "Thermovoltage Graph" (Fig. \ref{screen-inSThem}(b)) and "Temperature Graph" (Fig. \ref{screen-inSThem}(c)) windows. We start with the tip away from the heated sample ((i) in Fig. \ref{screen-inSThem}(b)(c)). Then the tip, at room temperature, makes nanoscale contact with the heated sample, and a temperature gradient is generated, which produces a thermoelectric voltage ((ii) in Fig. \ref{screen-inSThem}(b)). During the tip-sample nanocontact, the tip is slightly heated up by thermal conduction, causing a decrease in the thermovoltage after a few seconds. While the tip remains in nanoscale contact with the sample in (iii), the heater is turned off. As the temperature gradient begins to disappear, the thermoelectric voltage also decreases, demonstrating that the voltage is caused by the temperature difference between the tip and the sample. Note that the thermovoltage decreases at a faster rate than the temperature gradient. This is because the thermovoltage is measured at the nanocontact, where the tip and sample reach thermal equilibrium faster than in other regions.
\begin{figure*}
    \centering
     \includegraphics[scale=1]{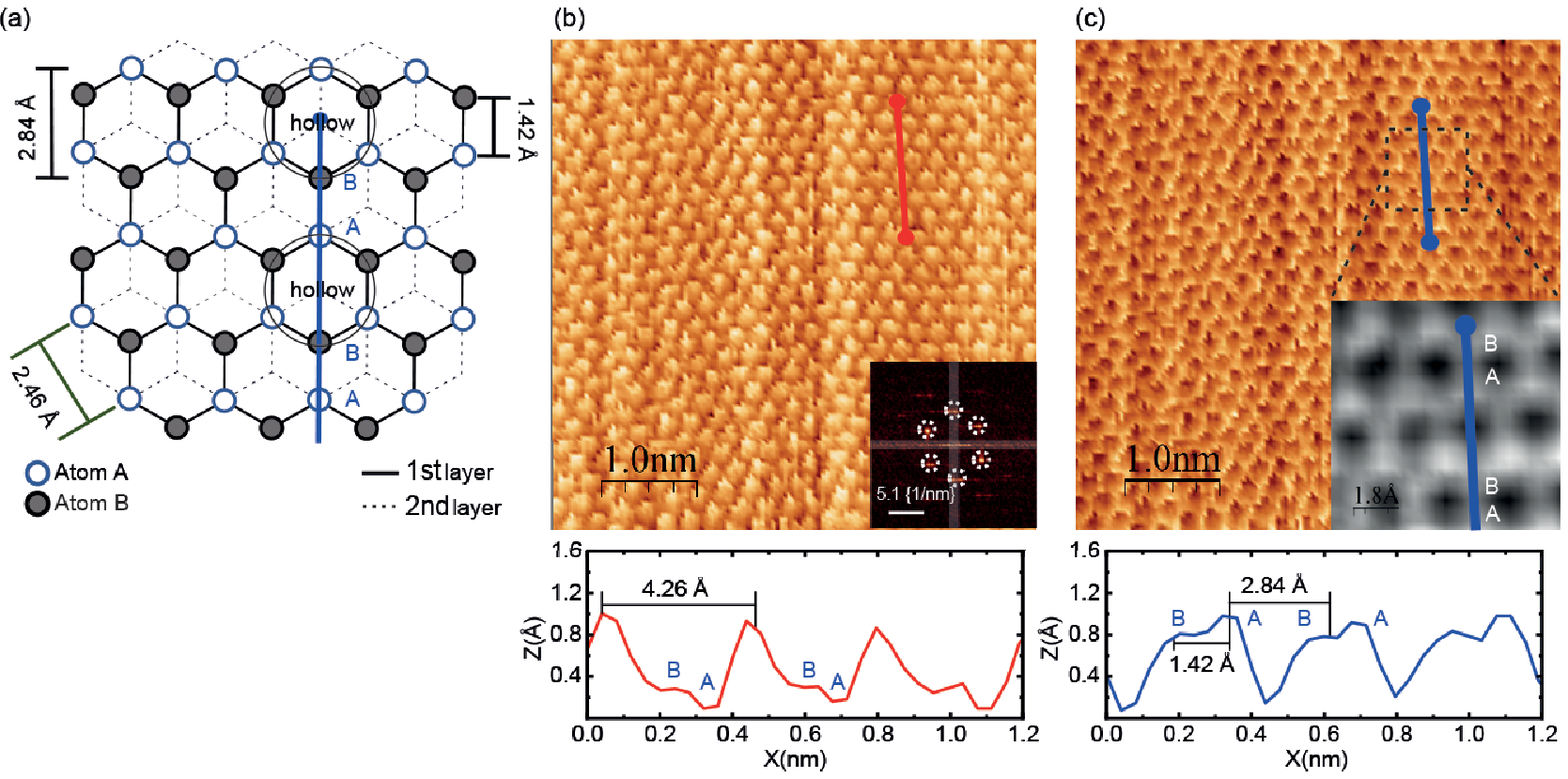}
    \caption{\label{STM-results}(a) Schematic representation of a typical STM image of bulk HOPG graphite using an Au tip. The distances between neighboring atoms are indicated, with 1.42 Å between adjacent atoms and 2.84 Å between opposite vertex atoms. The large solid circles in the image represent the size of an Au atom (2.9 Å in diameter), which coincides with the size of one hexagon in the graphite structure. Due to its comparable size, When the tip is positioned directly over the center of a hexagon, all six carbon atoms within that hexagon are simultaneously detected by the tip, resulting in a bright spot in the image at that location. This corresponds to the hollow-site lattice points in the HOPG structure. (b) STM image of HOPG graphite (taken at room temperature, $V_{bias} = 200mV$, $I = 1nA $), The bright spots in the image correspond to the locations where the tip is positioned over a hollow site (center of the hexagon), resulting in a strong tunneling current contribution from the six carbon atoms within that hexagon. The small image in the bottom corner shows the Fourier transform of the image, revealing the hexagonal lattice of the hollow sites. (c) Inverted Contrast Image Revealing the True Position of Carbon Atoms in a Honeycomb Lattice. The inset provides a magnified view of the region enclosed by the black square, highlighting a single atomic position. The profiles shown at the bottom of the image illustrate the height difference between the A and B atoms in the HOPG structure.}
\end{figure*}
\section{\label{vibrational-section}Electrical and Mechanical Noise Isolation}
In conventional STM operation, the resolution, scanning speed, and quality of topography/spectroscopy are often limited due to the interplay between the tunneling process and mechanical and electrical noise\cite{Ge2019}. In Fig. \ref{laboratory}, we show a picture of the STM/SThEM set-up. To minimize mechanical noise (vibration and acoustic) from the building, we have designed a home-built 1.5m x 1m soundproof box, made of brass and covered with soundproofing foam. The soundproof box (SB) is located on an optical table (VERE) with integrated passive vibration isolation. The STM/SThEM is mounted on a platform inside a home-built stainless steel high-vacuum (HV) chamber (Supplementary Fig. S1(a)). The vacuum chamber is placed inside the SB and suspended from the laboratory roof using vibration isolation springs and ropes. To reduce mechanical vibrations transmitted through the wires, we have carefully routed all of them through the soundproofing foam. Additionally, we have placed the most noise-sensitive instruments, including the STM electronic control, the I-V converter, and the voltage amplifier (DLPCA-200 by Femto\textregistered) to measure the tunnel current inside the soundproof box. The remaining instruments are positioned on an adjacent rack, as illustrated in Fig. \ref{laboratory}(a). 


The home-built vacuum chamber is capable of achieving a base pressure of $\sim 1\times10^{-5}$ mbar through a turbo molecular pump (TS85W1002) and measured with a Pirani gauge (RS-EV-D14642000). Additionally, the system incorporates a stainless steel glovebox (2800-2-B by Cleatech\textregistered) (Fig. \ref{laboratory}(b)), which provides a controlled atmosphere and clean environment that prevents humidity and oxidation during sample preparation when backfilled with nitrogen or argon. The STM/SThEM vacuum chamber is located inside the main chamber using a transfer chamber. We include a deterministic layer-transfer exfoliation stage (Fig. \ref{laboratory}(c)) through a 300x optical microscope and an XYZ manual stage, as described in Refs \citep{Castellanos-Gomez2014,Zhao2020,ibanes2022}. Inside the glovebox, we can exfoliate the sample or transfer layers to the STM sample holder in a clean atmosphere. The iris ports on the glovebox allow us to manipulate the STM/SThEM and access the vacuum chamber with bare hands while maintaining an inert atmosphere inside the glovebox.   


\begin{figure*}
    \centering
     \includegraphics[scale=1]{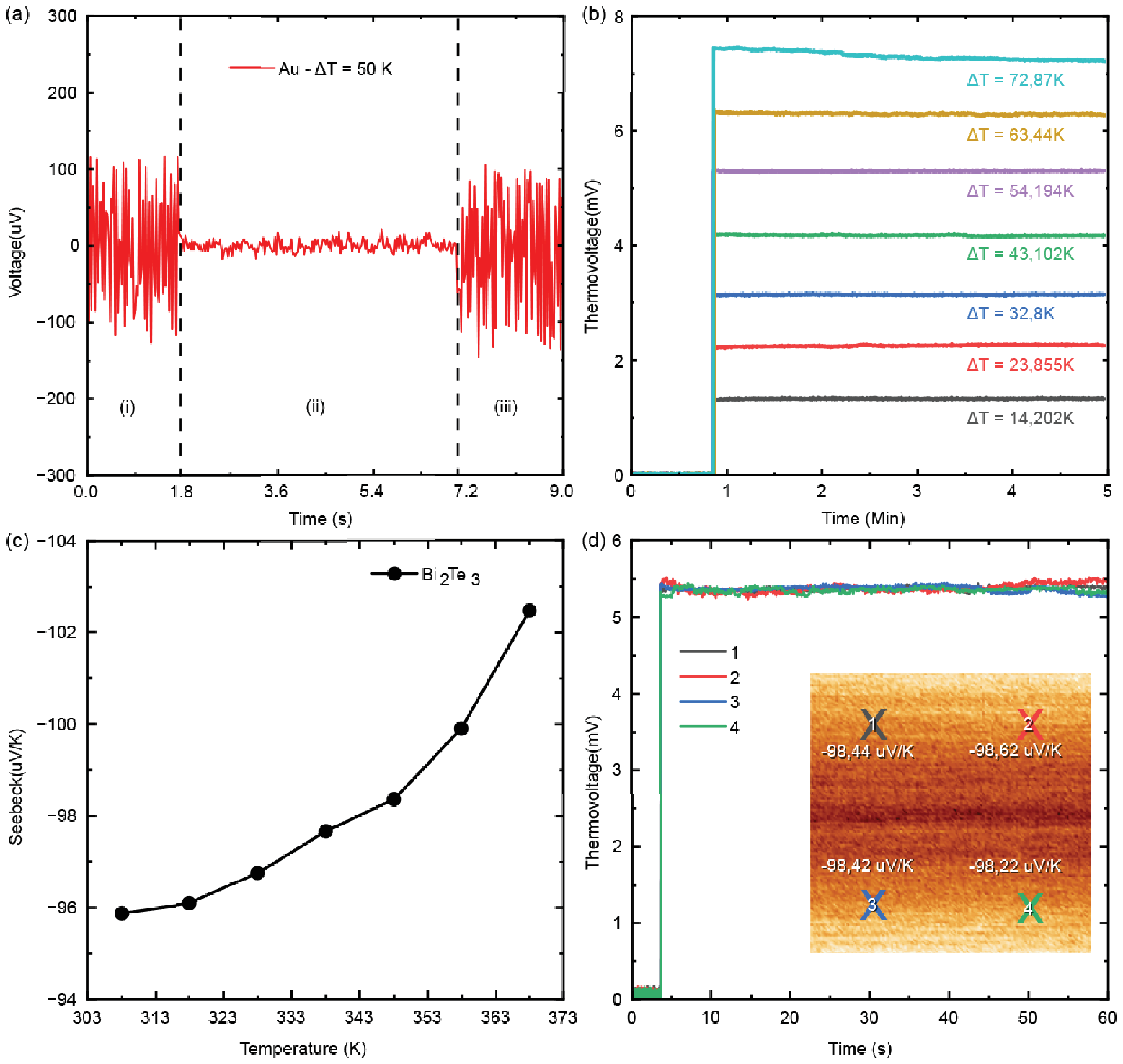}\par
    \caption{\label{SThEM_Resul} SThEM measurements: (a) Thermovoltage measurement on Au sample to demonstrate the behavior when the tip makes a nanoscale contact with a heated non-thermoelectric sample. In (i), the platinum-iridium tip is away from the heated sample. The tip made nanoscale contact with the heated sample at room temperature in (ii), and then retracted in (iii). (b) Induced thermoelectric voltage in $Bi_2Te_3$ due to different temperature gradients from 14.2K (black) to 72.87K (turquoise). (c) Seebeck coefficient of bulk $Bi_2Te_3$ sample versus temperature obtained from the curves in (b). (d) Thermoelectric voltage measured on four different zones of the sample surface as shown in the inset STM topographic image (size = 67 nm, Vbias = 350mV, I = 100pA).}
\end{figure*}

\section{\label{results-section}results}

Studies on the atomic resolution of graphite using STM can reveal important issues related to the microscopy technique itself. It is well-known that atomic resolution images on graphite only show the position of three out of six atoms in the same hexagon of the lattice (A and B atoms in Fig. \ref{STM-results}(a)) \cite{truegraphite}. Several theoretical and experimental studies attempt to explain this electronic asymmetry between neighboring carbon atoms\cite{Abraham1988,PhysRevLett.57.444,MATE1989473,Whangbo,PhysRevLett.57.444, PhysRevB.35.7790,GBinnig_1986,truegraphite,Ouseph1995,PhysRevB.36.4491,PhysRevLett.57.440}. Achieving atomic resolution in graphite (and in most materials) depends not only on the electronic properties of the sample but also on the coupling between the tip and the sample. 

Fig. \ref{STM-results}(b) shows atomic-resolution images of HOPG graphite in positive contrast. In this case, the bright spots do not correspond to individual atoms, but rather to the hollow centers of each hexagon of the lattice, creating a hexagonal lattice with a distance of 4.26Å between spots (see Fig. \ref{STM-results}(a)). This effect occurs because the radius of the atomically sharp tip (which is approximately 1.46Å for Au atoms) which is comparable to the radius of a hexagonal ring (1.42Å). When the tip is at the center of a hexagon, it receives tunneling current contributions from the six neighboring atoms, producing a bright spot at the center. However, by inverting the image contrast, it is possible to distinguish individual atoms, and the position of atoms A and B,  as shown in Fig. \ref{STM-results}(c).
Atoms A and B exhibit an asymmetry in their charge density. This is due to the fact that A atoms have neighbors directly below them in the second layer, whereas atoms B are located at the center of a hexagon in the second layer, with no atom immediately below them \cite{truegraphite,PhysRevB.37.8327}. This effect, resulting from electronic interactions between atoms of different layers, can be observed in the z-profile of figures \ref{STM-results} (bottom panels of b and c), where A atoms appear higher than B atoms. The observation of this small effect proves the high topographic resolution of our STM. 


To evaluate the SThEM mode, we conducted thermovoltage measurements using a Platinum-iridium tip, in both, a non-thermoelectric sample (Au) and a thermoelectric sample ($Bi_2Te_3$).  In Fig. \ref{SThEM_Resul}(a), we present the measurements on the Au sample. The sample was heated to 348K while the tip remained at room temperature. In (i), the tip was not in contact with the sample. We observed electrical noise around zero voltage induced by the sample heater. When the tip made contact with the sample in (ii), a noticeable reduction in electrical noise was observed. Although the noise level decreased, the thermovoltage response remained at zero. In (iii), the tip was moved away from the sample, similar to the configuration in (i). As expected, the noise level returned to its initial state. Consequently, in the absence of a thermoelectric response, the thermovoltage remains zero.

In Fig. \ref{SThEM_Resul}(b), a similar process was performed using a $Bi_2Te_3$ sample, a well-known thermoelectric compound with a thermopower ($S$) ranging from $-61 \mu V/K$ to $-170 \mu V/K$ near room temperature. The specific thermopower value depends on factors such as crystal structure, impurities, defects, or carrier concentration levels \cite{Zeipl2016, WU2013408, Goldsmid2014, Notes1986}. The $Bi_2Te_3$ sample was heated to different temperatures, ranging from 308K to 348K. Seven nanoscale contacts were made with the tip at room temperature at the same sample point for each temperature. The resulting thermovoltage curves are depicted in Fig. \ref{SThEM_Resul}(b) (from black to turquoise curves). To calculate the thermopower ($S$) using the formula $S = -\Delta(V)/\Delta(T)$, the maximum value of the thermovoltage at each nanoscale contact was utilized for each temperature (as shown in Fig. \ref{SThEM_Resul}(c)). For temperature differences below 70K, the thermovoltage response steadily increases until it reaches a maximum value, which then remains constant for several minutes. However, for temperature differences above 70K, a reduction in the thermovoltage is observed after a few seconds, which can be attributed to heat transfer to the tip. This behavior is evident in the turquoise curve in Fig. \ref{SThEM_Resul}(b), and it demonstrates that the thermovoltage response is a sensitive indicator of heat transfer processes, particularly for high-temperature differences ($\Delta T$).  

Fig. \ref{SThEM_Resul}(c) shows the calculated Seebeck coefficient as a function of the sample temperature for the measurements presented in Fig. \ref{SThEM_Resul}(b). As expected for $Bi_2Te_3$, the absolute value of the Seebeck coefficient increases with the sample temperature. To further validate the homogeneity and isotropy of S at the sample surface, we conducted four S measurements at different locations on a flat surface area, as shown in the inset STM image in Fig. \ref{SThEM_Resul}(d). The tests were performed consecutively at a sample temperature of 348K and $\Delta T=54.4K$, and all four measurements yielded a value close to $\bar{S}=98 \pm 0.03 \mu V/K  $. Notably, we did not observe any evidence of a reduction in thermovoltage due to heat transfer between measurements.

\section{Conclusion AND OUTLOOK}

In summary, we have presented the design, construction, and performance of a custom-built high-resolution scanning tunneling microscope operating at room temperature and pressure up to $1\times10^{-5}$ mbar. This setup enables atomic resolution imaging and achieves $\mu eV$ energy resolution. We have also detailed the modifications made to the STM to enable its operation as a Scanning Thermoelectric Microscopy. This allows for the simultaneous characterization of structural, electronic, and thermodynamic properties within a single technique. Furthermore, we discussed the integration of temperature control capabilities into the STM, enabling precise thermal manipulation of the sample. A user-friendly control interface was developed, allowing the entire system to be operated via USB standard communication through a computer. To ensure accurate measurements, we implemented various mechanisms to minimize noise stemming from mechanical vibrations and acoustic interference caused by the STM and SThEM operations. Lastly, we showcased the performance of the STM/SThEM system by demonstrating its ability to achieve atomic-resolution topographic imaging on monoatomic layers of Highly Oriented Pyrolytic Graphite (HOPG), as well as thermoelectric measurements on $Bi_2Te_3$ samples. Overall, the described STM/SThEM system provides a powerful platform for investigating nanoscale phenomena, offering a comprehensive range of capabilities for studying various materials and their properties.

\section*{Declaration of Competing Interest}
The authors declare that they have no known competing financial interests or personal relationships that could have appeared to influence the work reported in this paper.

\begin{acknowledgments}
The authors thank the support of the Ministerio de Ciencia, Tecnología e Innovación de Colombia (Grants No.122585271058), and the Cluster NBIC of Universidad Central (Colombia). E.H., and J.A.G. acknowledge the support of Departamento Administrativo de Ciencia, Tecnología e Innovación, COLCIENCIAS (Colombia) Convocatoria 784-2017. J.A.G. acknowledge the support of School of Engineering, Science, and Technology at Universidad del Rosario. P. G-G., K. V-B., and G. C-C., acknowledge the support of the School of Sciences and the Vice Presidency of Research Creation at Universidad de los Andes, and the financial support of FPIT-Fundación para la promoción de la investigación y la tecnología, of Banco de la Republica de Colombia, project number 4.687.  
\end{acknowledgments}



\bibliography
{References}

\end{document}